\documentclass{ws-procs975x65}

\begin{document}

\title{REPORT ON THE SESSION \\ LOOP QUANTUM GRAVITY: COSMOLOGY AND BLACK HOLES \\ OF THE 16TH MARCEL GROSSMANN MEETING}

\author{JORGE PULLIN and PARAMPREET SINGH}
\address{Department of Physics and Astronomy, Louisiana State
  University, Baton Rouge, LA 70803}



\begin{abstract}
We summarize the main results of 19 talks presented at the QG3 session (loop quantum gravity: cosmology and black holes) of the 16th Marcel Grossmann Meeting held online from July $5^{\mathrm{th}}$-10$^{\mathrm{th}}$, \,  2021.
\end{abstract}








\bodymatter

\vskip1cm

The QG3 session was organized in two parts with a total of 19 talks. The first part of the session was held on July 5th (9 talks) focussing on loop quantum gravity effects in black hole spacetimes, and the second part was on July 8th (10 talks) focussing on cosmological implications. A summary of talks in a chronological order is as follows. 

We-Cong Gan presented his joint work with Nilton O. Santos, Fu-Wen Shu, and Anzhong Wang on spherically symmetric polymer black holes \cite{Gan:2020dkb}. Their work is based on a model by Bodendorfer, Mele and M\"unch and depend on three independent parameters. An exploration of the parameter space was made showing a wealth of physical possibilities including singularity resolution, wormholes and space-times with singularities.

Harkirat Singh Sahota discussed joint work with Kinjalk Lochan \cite{sahota} about the quantization of the Lemaître--Tolman--Bondi model as a minisuperspace. It leads to a bounce eliminating the singularity and the talk explored how it manifests itself in the infrared regime to an observer looking at the collapse.

Aur\'elien Barrau discussed joint work with L\'eonard Ferdinand, Killian Martineau and Cyril Renevey \cite{Barrau:2021spy}. There have been several proposals discussed in the literature in which after their evaporation black holes tunnel into metastable white holes. They have been considered a dark matter candidate. The talk showed that such models are severely constrained and suggested some possible detection paths.

Parampreet Singh presented his work with Kristina Giesel and Bao-Fei Li on a non-singular collapse of a dust shell model in Lemaître--Tolman--Bondi spacetime \cite{Giesel:2021dug}. Two quantization prescriptions were studied. One using holonomies and triads, and another using holonomies and gauge-covariant fluxes. While both the prescriptions result in a bounce of the dust shell they lead to different mass gaps at which a trapped surface may form. Unlike in a quantization based on holonomies and triads, in presence of  gauge-covariant fluxes, a formation of white hole in the expanding phase when a black hole forms in the contracting phase is not guaranteed.  

Asier Alonso-Bardaji presented work with David Brizuela  where they consider holonomy and inverse-triad corrections in spherically symmetric models of loop quantum gravity \cite{Alonso-Bardaji:2021tvy}. It constructs systematically anomaly-free constraints with such corrections. It concludes that holonomy corrections are consistent in the presence of matter fields. Contrary to earlier claims, authors demonstrate a model which provides a family of deformed Hamiltonian constraints with scale-dependent holonomy corrections in the presence of a scalar matter field.

Cong Zhang talked about joint work with Yongge Ma, Shupeng Song and Xiangdong Zhang on the loop quantum gravity quantization of the interior of the Schwarzschild black hole \cite{Zhang:2021wex}. They study a Dirac observable corresponding to the ADM mass and note that zero is not in its spectrum, suggesting that a remnant is left after the evaporation of the black hole.

Alejandro Garc\'{\i}a Quasimodo talked about work with Guillermo Mena Marugán about some modifications to the Ashtekar--Olmedo--Singh black hole model, in particular the use of polymerization parameters that are Dirac observables \cite{Garcia-Quismondo:2021xdc}. This may modify the asymptotic behavior of the model. 

Saeed Rastgoo presented work with Keagan Blanchette, Saurya Das and  Samantha Hergott on loop quantum gravity corrections to the Raychaudhuri equation \cite{Blanchette:2020kkk}. These corrections imply defocusing of geodesics providing insights on the  eliminiation of singularities in this context.

Patrick Fraser analyzed how classical are the Gaussian states in loop quantum cosmology \cite{Fraser:2021ltu}. He showed that contrary to common intuition, such states do not saturate their uncertainty relations and there exist Gaussian states for which the fluctuations are arbitrarily large. However, the usual volume regularization procedure allows to suppress those fluctuations as much as one wishes.

V. Sreenath talked about alleviating tensions in the cosmic microwave background using Planck scale physics in a joint work with Abhay Ashtekar, Brajesh Gupt and Donghui Jeong \cite{Ashtekar:2021izi}. The idea is that there are several anomalies of low significance in the agreement of the predictions of the usual $\Lambda$-CDM model for the cosmic microwave background. Taken as a whole, however, they imply a tension between theory and data. The talk showed that considering loop quantum cosmology leads to a primordial power spectrum that is scale dependent at large scales. This can potentially alleviate the anomalies.

Sahil Saini discussed a matter-Ekpyrotic bounce model in loop quantum cosmology in a joint work with Bao-Fei Li and Parampreet Singh \cite{Li:2020pww}. Unlike previous attempts in this direction in loop quantum cosmology, Ekpyrotic phase was sourced by a potential motivated from Ekpyrotic scenarios. It was shown that in contrast to previous results where a fixed ultra-stiff equation of state was assumed,  the magnitude of power spectrum changes during evolution resulting from an Ekpyrotic potential. It was found that the bouncing regime only leaves imprints on the modes outside the scale-invariant regime. Further refinements in this direction are necessary since the spectral index shows inconsistency with the observational data.

Kristina Giesel presented a reduced phase space quantization for loop quantum cosmology in presence of inflationary potentials. This was a joint work with Bao-Fei Li and Parampreet Singh \cite{Giesel:2020raf}. Unlike the case of a  massless scalar field which has been addressed rigorously in loop quantum cosmology, singularity resolution at the quantum level in the presence of a potential has remained an open problem because of several conceptual and technical challenges. The idea is to introduce an additional clock degree of freedom to measure time evolution.  It was shown that singularity resolution occurs in the presence of potentials at the level of quantum difference equation. Some properties of non-singular solutions were discussed using effective dynamics.  

Guillermo Mena Marug\'an talked about his work with Beatriz Elizaga Navascu\'es and Rafael Jim\'enez-Llamas on the way potential effects primordial perturbations in kinetic dominated regimes in classical and loop quantum cosmology \cite{Navascues:2021brc}. Such a situation is of particular interest in loop quantum cosmology where so far kinetic dominated bounce has been studied. Authors found that the choice of Bunch-Davies vacuum may no longer be preferred in presence of such corrections. Further, there are changes in the effective mass in the Mukhanov-Sasaki equation.

Anzhong Wang discussed his joint work with Bao-Fei Li and Parampreet Singh on various subtleties in imposing initial conditions for cosmological perturbations in loop quantum cosmology and its modified versions (mLQC-I and mLQC-II) \cite{Li:2021mop}. The talk highlighted differences in choices which are forced by the assumptions in the dressed metric and hybrid approaches. Issues related to when to impose initial conditions -- whether at bounce or in the contracting phase, and constraints on possible choices were discussed. It was shown that a consistent choice of the initial conditions depends not only on the choice of a  potential but 
also on the two approaches used in loop quantum cosmology. 

Javier Olmedo talked about effects on primordial perturbations in a bouncing model in loop quantum cosmology for a Bianchi-I spacetime in a joint work with Ivan Agullo and V. Sreenath \cite{Agullo:2020wur}. They considered an evolution where anisotropies are non-vanishing at a bounce dominated by matter energy density.   Though anisotropies were diminished in the pre-inflationary branch, effects were found in the scalar and tensor power spectrum. Modification to the angular correlation functions and the way EB and TB correlations may arise was discussed.

Bao-Fei Li discussed features of primordial power spectrum in modified versions of loop quantum cosmology in a joint work with Javier Olmedo, Parampreet Singh and Anzhong Wang \cite{Li:2019qzr}. They discussed the way dressed metric and hybrid approaches to perturbations result in different signatures in the primordial power spectrum, especially for mLQC-I model. Differences with loop quantum cosmology in the infra-red and intermediate regimes were also found which show that different loop quantum prescriptions potentially leave distinct signatures in CMB.

Dimitrios Kranas talked about his joint work with Ivan Agullo and V. Sreenath on understanding the origin of various anomalies in the CMB using a bouncing model \cite{Agullo:2020fbw}. They find that  that non-Gaussian correlations between CMB modes and super-horizon wavelengths in the power spectrum can potentially explain a power suppression, a dipolar asymmetry, and existence of odd-parity correlations in the power spectrum. 

Adri\'a Delhom i Latorre discussed hos joint work with Gonzalo Olmo and Parampreet Singh \cite{adria}. They investigate the existence of a covariant action in $f(R)$ Palatini theories which can reproduce effective equations for modified versions of loop quantum cosmology. It was found that the Lagrangians which result in effective dynamics of LQC, mLQC-I and mLQC-II are part of a three-parameter family of f(R) theories, where two parameters are fixed by initial conditions at the bounce.

Hongguang Liu talked about his work with Muxin Han on a path integral approach based on reduced phase space loop quantum gravity to derive effective dynamical equations in which the graph changes dynamically in the physical time evolution \cite{Han:2021cwb}. They were able to derive the so-called improved dynamics from this method found earlier in loop quantum cosmology by Ashtekar, Pawlowski and Singh which results in a singularity resolution.  The method also generalizes the path integral formulation used in full LQG to take into account an additional real scalar field.

\section*{Acknowledgments}

This work was supported in part by grant NSF-PHY-1454832, NSF-PHY-1903799,  DFG-NSF-PHY-1912274, NSF-PHY-2110207 and funds of CCT-LSU and the Horace Hearne Institute for Theoretical
Physics.


\begin{thebibliography}{9}
\bibitem{Gan:2020dkb}
W.~C.~Gan, N.~O.~Santos, F.~W.~Shu and A.~Wang, "Properties of the spherically symmetric polymer black holes,”
Phys. Rev. D \textbf{102}, 124030 (2020)

\bibitem{sahota} H. S. Sahota, K. Lochan, {\it{To appear.}}

\bibitem{Barrau:2021spy}
A.~Barrau, L.~Ferdinand, K.~Martineau and C.~Renevey,
``Closer look at white hole remnants,''
Phys. Rev. D \textbf{103}, no.4, 043532 (2021)

\bibitem{Giesel:2021dug}
K.~Giesel, B.~F.~Li and P.~Singh, ``Non-singular quantum gravitational dynamics of an LTB dust shell model: the role of quantization prescriptions,"
[arXiv:2107.05797 [gr-qc]]

\bibitem{Alonso-Bardaji:2021tvy}
A.~Alonso-Bardaji and D.~Brizuela, ``Anomaly-free deformations of spherical general relativity coupled to matter,"
[arXiv:2106.07595 [gr-qc]]

\bibitem{Zhang:2021wex}
C.~Zhang, Y.~Ma, S.~Song and X.~Zhang, ``Loop quantum deparametrized Schwarzschild interior and discrete black hole mass,"
[arXiv:2107.10579 [gr-qc]]

\bibitem{Garcia-Quismondo:2021xdc}
A.~Garc\'\i{}a-Quismondo and G.~A.~Mena ~Marug\'an,
``Exploring alternatives to the Hamiltonian calculation of the Ashtekar-Olmedo-Singh black hole solution,''
Front. Astron. Space Sci. \textbf{8}, 701723 (2021)

\bibitem{Blanchette:2020kkk}
K.~Blanchette, S.~Das, S.~Hergott and S.~Rastgoo,
``Black hole singularity resolution via the modified Raychaudhuri equation in loop quantum gravity,''
Phys. Rev. D \textbf{103}, no.8, 084038 (2021)

\bibitem{Fraser:2021ltu}
P.~Fraser,
``Taming Fluctuations for Gaussian States in Loop Quantum Cosmology,''
Phys. Rev. D \textbf{103}, no.8, 086014 (2021)

\bibitem{Ashtekar:2021izi}
A.~Ashtekar, B.~Gupt, D.~Jeong and V.~Sreenath,
``Alleviating the Tension in the Cosmic Microwave Background using Planck-Scale Physics,''
Phys. Rev. Lett. \textbf{125}, no.5, 051302 (2020); A.~Ashtekar, B.~Gupt and V.~Sreenath, ``Cosmic Tango Between the Very Small and the Very Large: Addressing CMB Anomalies Through Loop Quantum Cosmology," Front. Astron. Space Sci. \textbf{8}, 76 (2021)


\bibitem{Li:2020pww}
B.~F.~Li, S.~Saini and P.~Singh,
``Primordial power spectrum from a matter-Ekpyrotic bounce scenario in loop quantum cosmology,''
Phys. Rev. D \textbf{103}, no.6, 066020 (2021)

\bibitem{Giesel:2020raf}
K.~Giesel, B.~F.~Li and P.~Singh,
``Towards a reduced phase space quantization in loop quantum cosmology with an inflationary potential,''
Phys. Rev. D \textbf{102}, no.12, 126024 (2020)

\bibitem{Navascues:2021brc}
B.~Elizaga~Navascu\'es, R.~Jim\'enez-Llamas and G.~A.~Mena ~Marug\'an,
``Primordial perturbations in kinetically dominated regimes of general relativity and hybrid quantum cosmology,''
[arXiv:2106.05628 [gr-qc]]

\bibitem{Li:2021mop}
B.~F.~Li, P.~Singh and A.~Wang,
``Phenomenological implications of modified loop cosmologies: an overview,''
Front. Astron. Space Sci. \textbf{8}, 701417 (2021)

\bibitem{Agullo:2020wur}
I.~Agullo, J.~Olmedo and V.~Sreenath,
``Predictions for the Cosmic Microwave Background from an Anisotropic Quantum Bounce,''
Phys. Rev. Lett. \textbf{124}, no.25, 251301 (2020)

\bibitem{Li:2019qzr}
B.~F.~Li, P.~Singh and A.~Wang,
``Primordial power spectrum from the dressed metric approach in loop cosmologies,''
Phys. Rev. D \textbf{101}, no.8, 086004 (2020); 
B.~F.~Li, J.~Olmedo, P.~Singh and A.~Wang,
``Primordial scalar power spectrum from the hybrid approach in loop cosmologies,''
Phys. Rev. D \textbf{102}, 126025 (2020)

\bibitem{Agullo:2020fbw}
I.~Agullo, D.~Kranas and V.~Sreenath,
``Anomalies in the CMB from a cosmic bounce,''
Gen. Rel. Grav. \textbf{53}, no.2, 17 (2021)


\bibitem{adria} A. D. i Latorre, G. J. Olmo, P. Singh, {\it{To appear.}}

\bibitem{Han:2021cwb}
M.~Han and H.~Liu,
``Loop quantum gravity on dynamical lattice and improved cosmological effective dynamics with inflaton,''
Phys. Rev. D \textbf{104}, no.2, 024011 (2021)
\end{thebibliography}
\end{document}